\documentclass[aps,prl,reprint, superscriptaddress]{revtex4-2}
\usepackage[english]{babel}

\usepackage{amsmath,amssymb,calrsfs}
\usepackage{graphicx}
\usepackage{subfigure}
\usepackage[colorlinks=true,citecolor=blue,urlcolor=blue,linkcolor = blue]{hyperref}  
\usepackage{dcolumn}
\usepackage{bm}
\usepackage{verbatim}
\usepackage{units}
\usepackage{upgreek}
\usepackage{multirow}
\usepackage{physics}
\usepackage{booktabs}
\usepackage[T1]{fontenc}
\usepackage[utf8]{inputenc}

\usepackage[dvipsnames]{xcolor}

\usepackage[normalem]{ulem}

\usepackage[none]{hyphenat}
\usepackage{float}

\newcommand{\Sec}[1]{{\textit{#1.---}}}

\newcommand{\TITLE}{Universal Behavior on the Relaxation Dynamics of Far-From-Equilibrium Quantum Fluids }

\begin{document}

\title{\TITLE }

\author{S. Sab }
\email[Corresponding author: ]{sarahsab@usp.br}
    \affiliation{Instituto de Física de São Carlos, Universidade de São Paulo, IFSC – USP, 13566-590, São Carlos, SP, Brasil}
    
\author{M. A. Moreno-Armijos}
	\affiliation{Instituto de Física de São Carlos, Universidade de São Paulo, IFSC – USP, 13566-590, São Carlos, SP, Brasil}

\author{A. D. Garc\'{i}a-Orozco}
	\affiliation{Instituto de Física de São Carlos, Universidade de São Paulo, IFSC – USP, 13566-590, São Carlos, SP, Brasil}

\author{G. V. Fernandes}
	\affiliation{Instituto de Física de São Carlos, Universidade de São Paulo, IFSC – USP, 13566-590, São Carlos, SP, Brasil}

\author{Y. Zhu}
    \affiliation{Université Côte d’Azur, CNRS, Institut de Physique de Nice (INPHYNI), 17 rue Julien Lauprêtre 06200 Nice, France}

\author{A. R. Fritsch}
	\affiliation{Instituto de Física de São Carlos, Universidade de São Paulo, IFSC – USP, 13566-590, São Carlos, SP, Brasil}

\author{H. Perrin}
    \affiliation{Université Sorbonne Paris Nord, Laboratoire de Physique des Lasers, CNRS UMR 7538, 99 av. J.-B. Clément, F-93430 Villetaneuse, France}

\author{S. Nazarenko}
    \affiliation{Université Côte d’Azur, CNRS, Institut de Physique de Nice (INPHYNI), 17 rue Julien Lauprêtre 06200 Nice, France}
    
\author{V. S. Bagnato}
	\affiliation{Instituto de Física de São Carlos, Universidade de São Paulo, IFSC – USP, 13566-590, São Carlos, SP, Brasil}
	\affiliation{Department of Biomedical Engineering, Texas A\&M University, College Station, Texas 77843, USA}	
\date{\today}

\begin{abstract}
Investigating the initial conditions that lead  many-body quantum systems to an out-of-equilibrium state is fundamental for understanding their thermalization dynamics.
In this work we observe the relaxation for two regimes of excitation that can drive the turbulent Bose-Einstein condensate into two distinct final states, and are defined by the amount of energy injected into the system. 
The subcritical regime is characterized by a lower injection of energy, which can lead to an inverse particle cascade and, consequently, to the BEC mode repopulation during the relaxation process. The supercritical regime is marked by a higher energy injection, that may lead to the BEC dissolution and a final thermal state.
In both cases we observe relaxation stages that exhibit the same key features:  a direct  cascade, a non-thermal fixed point with the same exponents, a prethermalization region and, finally, the thermalization of the system. In the final thermalization stage,  universal scaling is observed for both regimes, even though their final states are completely different.  By analyzing the coherence length of our turbulent cloud, we clearly visualize the recovery and the loss of the coherence for the subcritical and supercritical regimes after relaxation. These results indicate that the  evolution of turbulence occurs independent of its initial conditions and of the final state achieved.

\end{abstract}

\maketitle

Study of out-of-equilibrium quantum systems represents a challenging frontier in contemporary physics, as their dynamics do not match the usual properties of traditional thermal equilibrium \cite{eisert2015quantum, langen2015ultracold}. Investigating their evolution not only deepens our understanding of fundamental processes such as thermalization and  the build-up or loss of coherence, but also contributes to the development of quantum technologies \cite{chertkov2023characterizing, tatsuhiko2020general}, where control over nonequilibrium states is essential \cite{zhaoyu_2022}. 

The concept of universality provides a powerful tool for understanding the behavior of complex many-body systems. Near equilibrium, it allows identification of universality classes, where systems with different microscopic details display similar dynamical critical behavior \cite{hohenberg1977theory}.
Remarkably, universality can also emerge in out-of-equilibrium systems. An example is turbulence in quantum systems, where coherence effects enable the appearance of a far-from-equilibrium scaling manifesting itself as quantum turbulence \cite{seman2011route,madeira2020quantum,Henn2009}, as well as various forms of wave turbulence \cite{nazarenko2011wave, dossantosfilho2022incompressible, navon2016emergence, navon2019synthetic}. While turbulence is often analyzed in open systems, where external driving and dissipation sustain a stationary energy cascade and generate spatial scale invariance within an inertial range \cite{gatka2022emergence, karailiev2024observation}, universal dynamics can also arise in closed systems. In such cases, self-similar scaling in both space and time leads to the emergence of non-thermal fixed points (NTFPs)  \cite{pineiro2015universal, Schmied2019, mikheev2023universal, gazo2025science}, which act as attractor solutions for the far-from-equilibrium dynamics. Importantly, the scaling behavior near NTFPs is also captured by the framework of weak wave turbulence (WWT) theory, particularly in systems with high occupation and weak interactions. In this regime, the kinetic description provided by WWT yields scaling solutions to the wave kinetic equation (WKE) that match the scaling properties predicted near NTFPs.

Bose–Einstein condensates (BECs) provide a high degree of control over many different parameters and are therefore excellent platforms for studying out-of-equilibrium quantum systems.  The momentum distribution of  turbulent BECs  exhibits dynamical behavior and the emergence of a direct energy cascade \cite{seman2011route,navon2016emergence,garcia2022universal}.
The universal dynamical scaling of the turbulent BEC momentum distribution $n(k,t)$  during time evolution  takes the form 
\begin{equation}
    n(k,t)= \left(\frac{t}{t_{\mathrm{ref}}}\right)^\alpha n\left[ \left(\frac{t}{t_{\mathrm{ref}}}\right)^\beta k, t_{\mathrm{ref}}\right].
    \label{eq: Eq1}
\end{equation}
   Here  $k$ refers to the wavenumber,  $t$ is the hold time after  relaxation starts,  $t_{\mathrm{ref}}$  is an arbitrary reference time and $\alpha$ and $\beta$ are the universal exponents.
    The direct energy cascade results in a flux of particles  from low to high momenta, that leads to the depletion of the condensate mode \cite{garcia2022universal}.

Furthermore, the momentum distribution of an out-of-equilibrium quantum fluid can also exhibit an inverse particle cascade that is characterized by the self-similar solution of the second kind \cite{zhu2023direct,moreno2025observation,karailiev2024observation}  predicted by the WWT theory and is responsible for the condensate repopulation. The solution of the second kind follows the dynamical scaling of the form
\begin{equation}
    n(k,t)= \tau^\lambda n\left[ \tau^\mu k, t_{\mathrm{ref}}\right], \space \tau=(t^*-t)/(t^*-t_{\mathrm{ref}}),
    \label{eq: Eq2}
\end{equation}
where $t^*$   represents the time that the system achieves thermalization, and  $\lambda$ and $\mu$ are the universal exponents. As already observed, for the inverse particle cascade featuring BEC repopulation, $\lambda$ and $\mu$ are negative \cite{moreno2025observation, zhu2023direct}.

In this work, we study the thermalization processes of a three-dimensional BEC in a harmonic trap, driven into a turbulent state by a controlled excitation protocol. 
We observe that the thermalized state achieved after relaxation depends on the amount of energy injected into the system, leading either to BEC reconstitution or to a thermal state, referred to as dissolution, even though the mechanisms underlying turbulence establishment are independent of these final states.
Here we aim to extend the understanding of the thermalization stages by analyzing the case where it leads to the condensate dissolution.
\\

The experimental system has already been described in details  in past works \cite{seman2011route,garcia2022universal,moreno2025observation}. Briefly, we produce an atomic BEC of $^{87}$Rb in the hyperfine state $\ket{F=1,m_F=-1}$ confined in a magnetic Ioffe-Pritchard trap. The trap field is equivalent to an asymmetric harmonic oscillator characterized by axial and radial frequencies $\omega_x= 2\pi \times10.9(3)$ Hz and $\omega_r=2\pi \times90.3(5)$  Hz, respectively, producing a cigar-shaped condensate. At the end of radio-frequency evaporation, the atomic cloud reaches temperature $T\approx50$ nK with a condensate fraction $>80\%$ and total atom number $\sim 3\times 10^5$. After BEC production, an external magnetic field, produced by a set of coils in anti-Helmholtz configuration and misaligned with respect to the trap center, is turned on  to inject energy into the system.
The injected perturbation introduces oscillations to the atomic cloud in the original trap potential, creating rotations, deformations, and exciting collective modes. Once the excitation sequence ends, the external oscillating field is turned off and the system evolves in the trapping potential over a hold time $t$, which in this work ranges from 0 ms to 500 ms.
The excitation amplitude $A$ and the excitation time $t_\mathrm{exc}$ control the amount of energy injected into the atomic cloud.
Finally, the cloud is released from the trap and expands in time of flight $t_\mathrm{TOF}$ for 30 ms and an absorption image is taken along the $x-$axis.  From the absorption images we obtain the angular averaged momentum distribution $n(k,t)$ \cite{SupMat}. The total atom number is approximately conserved during the system evolution, except for experimental fluctuations, and the two-dimensional momentum distributions are normalized such that $2\pi \int k n(k,t)\ \mathrm{d}k=1$ \cite{SupMat}. Simulations with similar conditions to our system were already performed \cite{Holly2023}.
\\

Notably, the amount of energy injected into the system can lead to different final states. These states are determined by two excitation regimes, that we named \textbf{subcritical} and \textbf{supercritical}. The subcritical regime occurs when the energy injected is not sufficient to raise the temperature of the system above the critical temperature $T_c$ during relaxation. In this regime, the atomic cloud may exhibit an inverse particle cascade that leads to the repopulation of the BEC \cite{moreno2025observation, zhu2023direct, zhu2023self}. On the other hand, the supercritical regime occurs when the energy injected makes the temperature of the system go above $T_c$, which can lead to the condensate dissolution \cite{Temp_u0}. Important features of relaxation processes such as a direct cascade and a transient region were already observed in the subcritical regime \cite{moreno2025observation}.

\begin{figure*}[t]
\centering
     \includegraphics[scale=1]{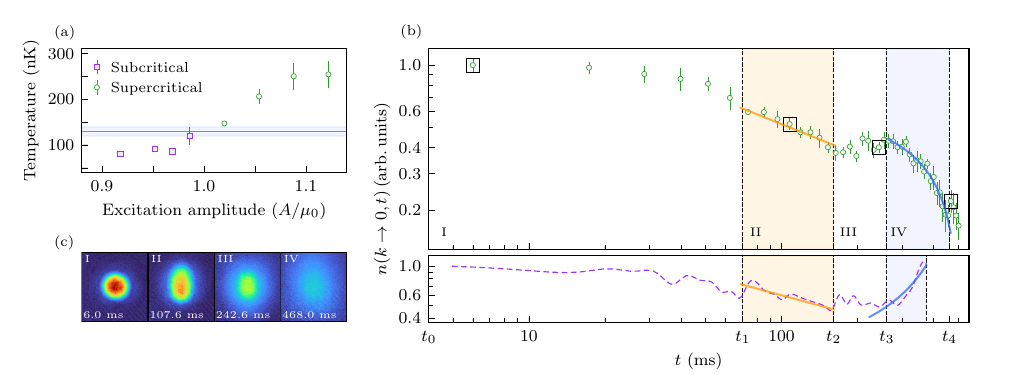}
    \caption{(a) Temperature measured at $t \approx 400$ ms for different excitation amplitudes. The horizontal blue line indicates the estimated critical temperature, and the shaded region denotes its standard deviation. For excitation amplitudes below (above) $\approx 1.00,\mu_0$, the system lies in the subcritical (supercritical) regime. (b) Time evolution of the  low-momentum population $n(k\rightarrow0,t)$, where $t_0$ indicates the end of the excitation stage. The green dots in the top panel correspond to the supercritical regime, while the dashed curve in the bottom panel represents the subcritical regime. Error bars denote one standard deviation \cite{SupMat}. The orange curves show the power-law fits $n(k\rightarrow0,t) \propto t^\alpha$ associated with the direct energy cascade. The blue curves show the power-law fits $n(k\rightarrow0,t) \propto (t^*-t)^\lambda$, corresponding to a second-kind self-similar solution that characterizes the BEC dissolution (top panel) and the BEC repopulation (bottom panel). (c) Absorption images taken during the supercritical regime for different holding times; each squared marker in panel (b) corresponds to one of these images.}
    \label{fig: Fig1}
\end{figure*}

Figure~\ref{fig: Fig1}(a) shows that, as the excitation amplitude increases, so does the temperature at $t \approx 400$ ms  \cite{TempAprox}, crossing the critical temperature that determines the limit between the subcritical and supercritical regimes. 

Observing the momentum distribution in the limit of low-momentum values $n(k\rightarrow 0, t)$ is very useful to identify the relaxation processes of the system \cite{madeira2024universal, moreno2025observation}, where the repopulation or dissolution of the BEC are observed.
Figure~\ref{fig: Fig1}(b) shows the time evolution of  $n(k\rightarrow 0, t)$ for two distinct regimes of excitation. The data in the top panel of Fig.~\ref{fig: Fig1}(b) correspond to amplitude $A=1.05 \mu_0$ in the supercritical regime, for which the different stages are described in more details below. On the other hand, the dashed curve  in the bottom panel of Fig.~\ref{fig: Fig1}(b) corresponds to amplitude $A=0.97 \mu_0$ illustrating the subcritical regime, which we can observe the presence of a direct cascade, a prethermalization stage and an inverse particle cascade \cite{moreno2025observation,zhu2023direct} indicating the BEC repopulation. Figure~\ref{fig: Fig1}(c) shows some illustrative absorption images that correspond to the relaxation stages of the supercritical regime, with  indices $I$, $II$, $III$ and $IV$.

 Equivalently to the subcritical case, during the hold time we observe different relaxation stages in $n(k\rightarrow0,t)$ for the supercritical regime. From $t_0=0$ ms to $t_1=70$ ms we have the start of  relaxation in which we observe a slow depopulation of $n(k\rightarrow0,t)$.
In the sequence, from $t_1=70$ ms to $t_2=160$ ms, we observe a linear decrease of the low-momentum population in log-log scale, characteristic of the establishment of turbulence and displaying a direct energy cascade \cite{garcia2022universal}. In this time interval the system exhibits universal scaling behavior and the low-momentum distribution obeys  the power law $n(k\rightarrow0,t) \propto t^\alpha $ \cite{madeira2024universal}. Therefore, we fit the data to the power law and obtain $\alpha=-0.50(7)$. The fitting, represented by the orange line, is shown in Fig.~\ref{fig: Fig1}(b).

We can also observe a transient region from $t_2=160$ ms to $t_3=260$ ms, which may correspond to a prethermalization of the system  where $n(k \rightarrow 0, t)$ is approximately constant \cite{Smith_2013,langen2016prethermalization}. This quasi-stationary region is followed by a final stage, from $t_3=260$ ms to $t_4=460$ ms, where $n(k\rightarrow 0,t)$ decreases over time as the system depopulates the BEC achieving a final thermalized state.

Inspired by the inverse particle cascade case \cite{moreno2025observation}, we try to describe the dissolution stage as a self-similar behavior of the second kind characterized by Eq.~(\ref{eq: Eq2}), since both exhibit similar behaviors in the same time interval, but with opposite particle flux direction.
In this sense, this self-similar behavior of the second kind  might display the low-momentum limit $n(k\rightarrow0,t) \propto (t^*-t)^\lambda$. By fitting our data we obtain $\lambda = 0.6(1)$ and $t^*=520(10)$ ms indicating that the dissolution may be described by Eq.~(\ref{eq: Eq2}), but with positive exponents.  The fitting, represented by the blue curve is shown Fig.~\ref{fig: Fig1}(b).

\begin{figure*}[t]
\centering
     \includegraphics[width=\textwidth]{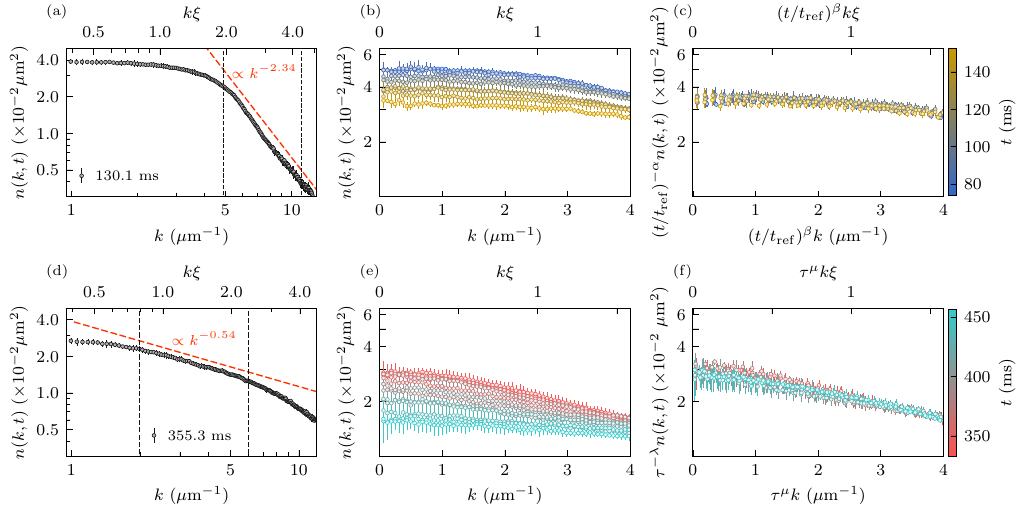}
    \caption{ Momentum distributions $n(k,t)$ for different relaxation stages of the supercritical case. (a)-(c) are data from $t_2$ to $t_3$ where we observe a direct cascade, while (d) to (f) are the data from the thermalization stage. In (a) the momentum distribution for $t = 130.1$ ms shows a direct cascade with a power law $k^\gamma$ with $\gamma=-2.34(2)$, represented by the the red dashed line. The vertical dashed lines represent the momentum region where we made the fitting.
    In the termalization stage, shown in (d) for $t = 355.3$ ms, we can also observe a power law $k^{-\nu}$ from which we obtain $\nu=0.54(1)$, the dashed vertical lines represent the region where we made the fitting. In (b) we show the momentum distribution for different hold times and (c) its respective scaling following Eq.~(\ref{eq: Eq1}),  with universal exponents $\alpha = -0.57(9)$ and $\beta = -0.25(8)$. In (e) we show the momentum distribution for different hold times and  its scaling following Eq.~(\ref{eq: Eq2}), with $\lambda=0.6(2)$ and $\mu = 0.8(3)$. The error bars correspond to the standard deviations \cite{SupMat}.}
    \label{fig: Fig2}
\end{figure*}

 The WWT theory can well describe our turbulent system considering interactions via weakly nonlinear processes between density waves, that were introduced in our system by the excitation protocol. Particularly, the WWT theory describes the four-wave mixing process via the WKE, which predicts self-similar solutions \cite{zakharov1992kolmogorov,nazarenko2011wave}.

Analyzing the full momentum distributions $n(k,t)$, we can identify the power law dependence in an inertial range. For the direct energy cascade, identifying the power law  $n(k) \propto k^{-\gamma}$ is fundamental to establish the type of turbulence that is evolving in our system.  Relevantly, the Kolmogorov-Zakharov stationary solution of the WKE gives $n(k) \propto k^{-2}$ for the direct energy cascade, neglecting logarithmic corrections. Analyzing the momentum distribution for a hold time in the interval that the direct cascade emerges, fitting by the power law distribution results in  $\gamma=2.34(2)$, as shown in Fig. \ref{fig: Fig2}(a). 
Furthermore, the direct energy cascade, described by the self-similar solution of first kind, exhibits universal dynamical scaling in the form of Eq.~(\ref{eq: Eq1}). The WWT theory predicts $\alpha=-2/3$ and $\beta=-1/6$ universal exponents for an infinite and homogeneous potential   \cite{zhu2023direct,zhu2023self}. In our experiment  we obtained $\alpha=-0.57(9)$ and $\beta=-0.25(8)$, applying the same method as in Refs.~\cite{garcia2022universal,moreno2025observation}. The discrepancies to the theoretical values are attributed to the fact that we have an inhomogeneous potential and that we only have access to the projected momentum distribution of our system. By using Eq.~(\ref{eq: Eq1}) with the experimental parameters $\alpha $ and $\beta$, we can observe the collapse of momentum distributions in the direct cascade region, shown in Figs.~\ref{fig: Fig2}(b,c).

These results suggest that the mechanisms responsible for turbulence establishment are the same as those observed at lower amplitudes (subcritical regime) that lead to the condensate repopulation \cite{moreno2025observation}. 
In Tab.~\ref{tab:table1} we show a comparison of the universal exponents of the direct cascade for both regimes.

For the self-similar behavior of the second kind, it is also important to identify presence of a power law in the momentum distribution $n(k) \propto k^{-\nu}$. 
As indicated in Fig.~\ref{fig: Fig2}(d) we observe a power law with $\nu=0.54(1)$ that describes well the behavior of our experimental data in the interval between $\sim2$ $\mu$m$^{-1}$  and  $\sim6$ $\mu$m$^{-1}$.
Subsequently, from WWT we have that $\lambda = \nu/(2-2\nu) \approx 0.59$. Since $n(k)$ is a decreasing function, it follows that $\mathrm{d}n(k)/\mathrm{d}k <0$, implying that  $\nu$ is positive in the dissolution interval. Knowing that for the emergence of the inverse cascade, as the BEC repopulation, $\lambda = \nu/(2-2\nu) <0$, we must have $\nu >1$. In a similar way, for the dissolution of the BEC, we have $\lambda = \nu/(2-2\nu) >0$, such that $0<\nu<1$.

The momentum distributions during the self-similar behavior of the second kind also collapse into a universal function, as shown in Figs.~\ref{fig: Fig2}(e,f), in which we  applied the same method used for the first direct particle cascade, now using the dynamical scaling given by Eq.~(\ref{eq: Eq2}). From our experimental data we obtained the exponents $\lambda=0.6(2)$ and $\mu=0.8(3)$, fixing $t^*=520$  ms. 
Notably, the dissolution at the supercritical regime seems to fit the  same  description used for inverse particle cascade at the subcritical regime, also described by Eq.~(\ref{eq: Eq2}). These results evidence that the two regimes evolve in a similar manner, exhibiting both the direct cascade and the prethermalization stage, differing only in their dynamics during the final stage. Table.~\ref{tab:table1} shows the universal exponents of the thermalization stage for both regimes.

\begin{table*}[]
\resizebox{0.8\textwidth}{!}{%
\begin{tabular}{l|cccccc|cccccc|}
\multirow{2}{*}{}                                                                                                   & \multicolumn{6}{c|}{\textbf{Theoretical} Refs.~\cite{zhu2023direct,zhu2023self}}                                                                                                                                                                                    & \multicolumn{6}{c|}{\textbf{Experimental}}                                                                                                                            \\ \cline{2-13} 
                                                                                                                    & \multicolumn{3}{c|}{Direct cascade}                                                                                                                    & \multicolumn{3}{c|}{2nd kind}                                       & \multicolumn{3}{c|}{Direct cascade}                                                           & \multicolumn{3}{c|}{2nd kind}                                         \\ \hline
                                                                                                                    & \multicolumn{1}{c|}{$\gamma$}            & \multicolumn{1}{c|}{$\alpha$}                        & \multicolumn{1}{c|}{$\beta$}                         & \multicolumn{1}{c|}{$\nu$} & \multicolumn{1}{c|}{$\lambda$} & $\mu$ & \multicolumn{1}{c|}{$\gamma$} & \multicolumn{1}{c|}{$\alpha$} & \multicolumn{1}{c|}{$\beta$}  & \multicolumn{1}{c|}{$\nu$} & \multicolumn{1}{c|}{$\lambda$} & $\mu$   \\ \hline
\begin{tabular}[c]{@{}l@{}}Subcritical regime\\ Ref.~\cite{moreno2025observation}\end{tabular} & \multicolumn{1}{c|}{\multirow{2}{*}{-2}} & \multicolumn{1}{c|}{\multirow{2}{*}{-2/3}} & \multicolumn{1}{c|}{\multirow{2}{*}{-1/6}} & \multicolumn{1}{c|}{1.52}  & \multicolumn{1}{c|}{-1.46}     & -1.04 & \multicolumn{1}{c|}{-2.4}     & \multicolumn{1}{c|}{-0.5(1)}  & \multicolumn{1}{c|}{-0.25(7)} & \multicolumn{1}{c|}{1.6}   & \multicolumn{1}{c|}{-1.5(5)}   & -0.9(3) \\ \cline{1-1} \cline{5-13} 
Supercritical regime                                                                                                & \multicolumn{1}{c|}{}                    & \multicolumn{1}{c|}{}                                & \multicolumn{1}{c|}{}                                & \multicolumn{3}{c|}{-}                                              & \multicolumn{1}{c|}{-2.3(1)}     & \multicolumn{1}{c|}{-0.55(5)} & \multicolumn{1}{c|}{-0.23(3)} & \multicolumn{1}{c|}{-0.55(1)}  & \multicolumn{1}{c|}{0.64(4)}    & 0.8(2)  \\ \hline
\end{tabular}%
}
\caption{Power laws and universal exponents for the subcritical \cite{moreno2025observation} and supercritical regimes.  The exponents $\gamma$, $\alpha$ and $\beta$  are related to the direct particle cascade, and $\nu$, $\lambda$ and $\mu$ to the self-similar solution of the second kind. The theoretical results correspond to a homogeneous system. The experimental exponents for the supercritical were computed by averaging the exponents obtained from the four excitation amplitudes corresponding to this regime in Figure~\ref{fig: Fig1}(a).}
\label{tab:table1}
\end{table*}

\begin{figure}
    \centering
    \includegraphics[width=1.0\linewidth]{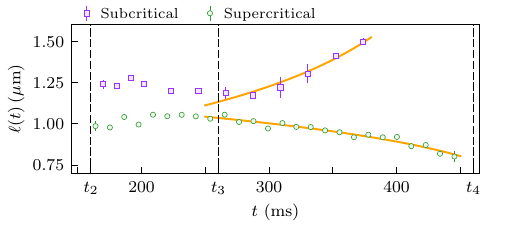}
    \caption{Coherent length as a function of the hold time for subcritical and supercritical regimes. In interval from $t_2$ to $t_3$, corresponding to the prethermalization stage, $\ell(t)$ is approximately constant. From $t_3$ to $t_4$ we have the thermalization stage. The orange curves represent the fitting of the power law $\ell(t) \propto (t^*-t)^{\lambda'}$, from which we  obtained $\lambda^{\prime}_{\mathrm{sub}}=-0.49(5)$  for the recovery of coherence and $\lambda^{\prime}_{\mathrm{sup}}=0.19(1)$ for the continued loss of coherence in subcritical and supercritical regimes, respectively. The dashed vertical lines represents $t_2$, $t_3$ and $t_4$. The error bars correspond to the standard deviations \cite{SupMat}.}
    \label{fig: Fig3}
\end{figure}

For our measurements it is also interesting to quantify the loss or recovery of the coherence by computing the coherence length during the relaxation of the  out-of-equilibrium quantum fluid \cite{bloch2000measurement, martirosyan2025universal}. Relevantly, the coherence length $\ell(t)$ is obtained from the first-order correlation function $g_1(r)$ decay that is related to the momentum distribution $n(k,t)$ by its inverse Fourier transform \cite{SupMat} \cite{Bray01061994,naraschewski1999}. In this sense, from the two-dimensional momentum distributions obtained from time-of-flight absorption images we could determine the coherence length during the system's time evolution, for both subcritical and supercritical regimes. The results are shown in Fig.~\ref{fig: Fig3}. Our observations focus solely in the last two time intervals of the relaxation stages.
From $t_2$ to $t_3$, we have the prethermalization of the system, where the coherence length remains approximately constant.
Subsequently, from $t_3$ to $t_4$ corresponding to the system final thermalization, we could observe that $\ell(t)$ evolves following the power law $(t^*-t)^{\lambda^{\prime}}$  for both regimes, with $\lambda^{\prime}_{\mathrm{sub}}=-0.49(5)$ representing the  recovery of coherence for the subcritical case and $\lambda^{\prime}_{\mathrm{sup}}=0.19(1)$ referent to the continued loss of the coherence for the supercritical case. The overall coherence length in the subcritical regime is always higher than the values obtained for the supercritical regime.
\\

In conclusion, our work shows that the final stage of the evolution of a quantum system driven out of equilibrium can lead either to a new degenerate quantum state (subcritical regime) or to a thermal state (supercritical regime),  which is expected. We observe that the final state is controlled by the amount of injected energy, but the route the system takes to thermalization shows similar features, such as the relaxation time in each stage. The direct energy cascade and NTFPs indicate the establishment of turbulence, exhibiting universal dynamical scaling. Notably, the universal exponents $(\alpha, \beta)$ are similar for both regimes, showing that the system’s scalability belongs to the same universality class, independent of the initial or final conditions.  The presence of the quasi-stationary region in low momentum distribution indicates the presence of a prethermalization stage. Finally, the system achieves the final thermalization stage, that may lead to the BEC reconstruction, characterized by an inverse particle cascade, or to dissolution of the BEC characterized by a self-similar behavior of the second kind resulting in a thermal cloud. In both cases, the thermalization to the final state displays a universal behavior that obeys Eq.~(\ref{eq: Eq2}), with the exponents being negative for the recovery of the condensate mode, and positive for its dissolution. Analyzing the coherence length we could clearly observe the recovery and loss of coherence expected for each regime.
\\

The authors thank M.A.~Caracanhas for fruitful discussions. This work was supported by the São Paulo Research Foundation (FAPESP) under grants 2013/07276-1, 2014/50857-8, by the cooperation between FAPESP and the French National Research Agency (ANR) through project RELAQS number 2024/04637-8 and ANR-24-CE30-6525-01, and by the ANR through project VORTECS (Grant No. ANR-22-CE30-0011). The authors also acknowledge the support of the National Council for Scientific and Technological Development (CNPq) under the grants 465360/2014-9. This work was granted access to the high-performance computing facilities under the OPAL infrastructure of Université Côte d’Azur, supported by the French government through the UCAJEDI Investments in the Future project, managed by the National Research Agency (ANR), under Reference No. ANR-15IDEX-01. S.S, M.A.M-A, A.D.G-O, and A.R.F. acknowledge the support from FAPESP -- Finance codes No. 2024/14764-7, No. 2025/13137-1, No. 2025/07547-2, No. 2024/08433-8 and No. 2024/21658-9. V.S.B. acknowledges Texas A\&M University at College Station - TX.

\clearpage
\pagebreak
\twocolumngrid
\begin{center}
\textbf{\large Supplemental Material: \TITLE}
\end{center}
%%%%%%%%%% Prefix a "S" to all equations, figures, tables and reset the counter %%%%%%%%%%
\setcounter{equation}{0}
\setcounter{figure}{0}
\setcounter{table}{0}
\makeatletter
\renewcommand{\theequation}{S\arabic{equation}}
\renewcommand{\thefigure}{S\arabic{figure}}

\Sec{Excitation protocol}
Our excitation setup responsible for taking the system out of equilibrium is composed of a pair of coils in anti-Helmholtz configuration with one of the coils misaligned in $z$ and $y$ directions with respect to the symmetry axis of the trapping potential.
 By applying an oscillatory electric current through the coils, we generate a time-oscillating gradient that shifts and compresses the trapping potential center, inducing rotations and compressions in the atomic cloud. Even though our trap is symmetric in $y$ and $z$, the excitation couples energy predominantly to the $z$ direction, since we observe momentum spread mostly in this direction. To explore different excitation regimes, we can vary the magnitude, frequency, phase and the duration of the electric current that passes through the coils. In this work we fixed the excitation period $\tau$, excitation time $t_{\mathrm{exc}}=7\tau$ and excitation frequency $\omega_{\mathrm{exc}}= 2\pi/\tau=2\pi \times110$ Hz for all the measurements, the only variable parameter is the magnitude. In our system, the magnitude represents the excitation amplitude that is proportional to the peak voltage,  $V_{\mathrm{pp}}$, measured in a resistor connected in series with the coils.. In this sense, we calibrated the potential generated by the excitation coils in terms of the chemical potential of the center of the cloud in equilibrium, $\mu_0$, as a reference parameter of our system.
 For the calibration, we pulse the excitation field with a fixed current immediately after the cloud is released from the trap. The force applied in the cloud can be cast as
\begin{equation}
    F=m \frac{dv}{dt}=-\frac{dU}{dz} \approx -\frac{\Delta U}{\Delta z}
    \label{eq: EqS1}
\end{equation}
where $\Delta z$ is the cloud size along the z direction, $t$ is the time of the pulse, $m$ is the atom mass, $v$ is the velocity of the cloud after the pulse and $U$ the potential of the force. As the measurement is in time of flight, the Thomas-Fermi radius in $z$ increases as the cloud expands during the pulse, we consider a time-varying size, or $\Delta z =2R_{z}(t)$. Note that the pulse duration $\Delta t $ is much smaller than the time of flight $ t_{\mathrm{ToF}}$, so we can estimate the velocity increase due to the pulse as $dv \approx\Delta v = d/t_{\mathrm{ToF}}$, with $d$ being the  displacement of the center of mass in $z$. In this sense we have the injected energy
\begin{equation}
    \Delta U= \frac{2md}{t_{\mathrm{ToF}}} \frac{1}{\int_{t_i}^{t_f}[R_z(t)]^{-1}dt}
    \label{eq: EqS2}
\end{equation}
where $t_i$ and $t_f$ are the instants when the pulse starts and stops respectively. As we have an axially symmetric cigar shaped trap with respect to $x$, we consider the Castin-Dum approximation $R_z(t) \approx R_0 \sqrt{1+\lambda(t)^2}$ \cite{castin_dum1996}, where $R_0$ is the Thomas-Fermi radius in situ and $\lambda(t)=\omega_r t$, with $\omega_r = (\omega_y \omega_z)^{1/2} \approx\omega_z$ such as $\omega_y \approx \omega_z$. The potential variation $\Delta U$, expressed in terms of $\mu_0$ is what we define as the excitation amplitude $A$. From our experiment conditions, we have that $1$ $V_{\mathrm{pp}}$ $\approx$ $0.34\mu_0$.

\begin{figure}
    \centering
    \includegraphics[width=\linewidth]{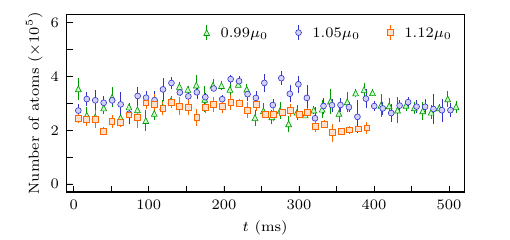}
    \caption{Total number of atoms  during hold time for three amplitudes. The number is approximately constant over time, with small experimental fluctuations. The error bars correspond to the standard deviation of the mean value of all measurements.}
    \label{fig:FigS1}
\end{figure}

 In this work we monitored the number of atoms  over time to ensure that we have a closed system with an approximately constant number of particles. Figure~\ref{fig:FigS1} shows the number of atoms for three excitation amplitudes during hold time. 
The number remains approximately constant with small deviations due to experiment shot-to-shot fluctuations. The measurement was repeated at least six times for each hold time $t$. The error bars represent the standard deviation of the mean value of all measurements.
\\

\Sec{Measurement of the momentum distributions}
We obtain the \textit{in situ} time-dependent momentum distributions $ \tilde n(k,t)$ of our atomic cloud via absorption images after the time of flight. The two-dimensional (2D) image is a projection of the three-dimensional (3D) density of the system, from which we extract the column density in the $yz$ plane. Our position coordinates $r$ of the atomic distribution are converted to momentum coordinates $k$, such as $r=\hbar t_{\mathrm{ToF}} k/m$ and $k=(k_y^2+k_z^2)^{1/2}$. In this work we normalize the 2D momentum distribution by the number of atoms, $n(k,t)=\tilde n(k,t)/N(t)$, to avoid uncertainties from small fluctuations. For our measurements  we release the atoms from the harmonic trap at hold times multiple of the radial oscillation period $T_r$ when the in situ atomic cloud passes through the trap's center and the atomic distribution reflects the momentum distribution, which is amplified by time of flight.

\begin{figure*}
    \centering
    \includegraphics[width=\linewidth]{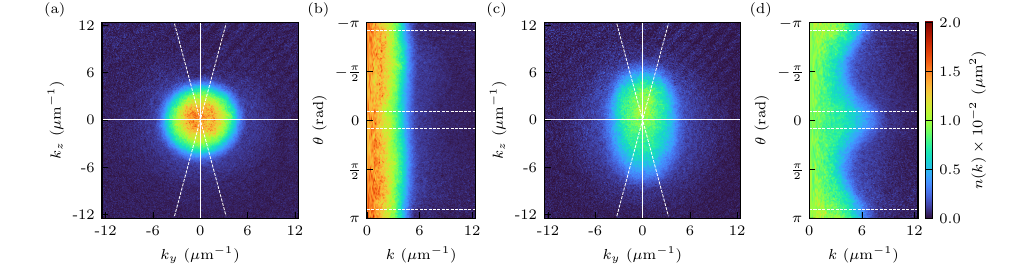}
    \caption{Single shot illustrations of the angular averaging procedure to compute the momentum distributions for the excitation amplitude $A=1.054\mu_0$ at hold times (a) $t=6.03$ ms and (c) $t=107.6$ ms. In both cases, the dashed lines delimit two $30^{\circ}$ angle regions centered around the major axis of expansion of the cloud, where $n(k)$ is computed. Images (b) and (d) represent the same information as (a) and (c) respectively, but in polar coordinates. }
    \label{fig:FigS2}
\end{figure*}

For an isotropic 2D distribution $n(k_y,k_z)$, we extract the radial momentum distribution by doing the angular average. However, when we take the BEC to an out-of-equilibrium state, we observe that the momentum spreads mainly in $z$ direction and the 2D distribution turns anisotropic. To minimize the anisotropy effects, when calculating the momentum distribution, we take the angular average restricted to an aperture of $30^{\circ}$ around the $z$-axis. Two examples in cartesian coordinates  for the excitation amplitude $A=1.05\mu_0$ at holding times $t=6.03$ ms and $t=107.6$ ms are illustrated in Fig.~\ref{fig:FigS2}(a) and Fig.~\ref{fig:FigS2}(b), respectively.  Their corresponding 2D distributions in polar coordinates are shown in Fig.~\ref{fig:FigS2}(c) and Fig.~\ref{fig:FigS2}(d), the dashed lines limit the angular regions in $\theta$ where we take the average.  From the momentum distributions $n(k,t)$ we obtain the values of  $n(k\rightarrow 0,t)$ by taking the average of $n(k,t)$ for $k\le 0.33$ $\mu \mathrm{m}^{-1}$. The pixel size in our absorption images corresponds to momentum $0.066$ $\mu\mathrm{m}^{-1}$.

 \begin{figure}
    \centering
    \includegraphics[width=\linewidth]{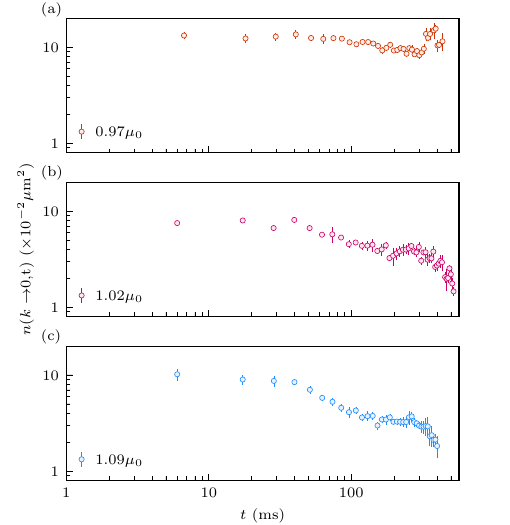}
    \caption{Time evolution of $n(k \rightarrow0, t)$ for the excitation amplitudes $ A = 0.97\mu_0$, $1.02\mu_0$ and $1.07\mu_0$, panels (a)-(c), respectively}
    \label{fig:FigS3}
\end{figure}

The observation of the dissolution of the atomic cloud into a thermal state depends on the amount of the energy injected into the system. If not enough energy is injected, the system can not even establish turbulence.
Figure~\ref{fig:FigS3} shows $n(k\rightarrow0,t)$ for three distinct amplitudes of $A=0.97\mu_0$, $1.02\mu_0$ and $1.07\mu_0$. For $A=0.97\mu_0$, the energy injection is low but enough so that we can observe the appearance of an inverse particle cascade around $t\approx300$ ms. For  amplitude  $1.02 \mu_0$ we can already observe the emergence of the dissolution of the system, around $t\approx 400$ ms. On other hand, for $1.07\mu_0$, we can observe the beginning of the dissolution, but could not observe the entire thermalization due to dilution of the atomic cloud.
\\

\Sec{Computing the coherence length from 2D momentum distributions}
To  determine the coherence length of our system, we must compute the first order correlation function $G_1(\vec r,\vec{r}^\prime)=\braket{\psi^*(\vec r)}{\psi(\vec{r}^\prime)}$, where $ \psi$ represents the wavefunction that describe the atomic cloud \cite{PhysRevA.59.4595}. From our measurements we have direct access to the 2D momentum distribution, that can be defined as the Fourier transform $n(\vec k)= \mathcal{F}_{2D}[\braket{\psi^*(\vec r)}{\psi(\vec{r}^\prime)}]$. We can relate the first order correlation function for a non-uniform gas with the 2D inverse Fourier transform of the momentum distribution as follows

\begin{equation}
    \bar G_1(\Delta \vec{r})= \mathcal{F}_{2D}^{-1}[n(\vec k)] = \frac{1}{S} \int d^2\vec{k} \, e^{i \vec{k}\cdot \Delta\vec{r}} n(\vec{k}),
    \label{eq: EqS3}
\end{equation}
with
\begin{equation}
    \bar{G}_1(\Delta \vec{r}) = \frac{1}{S} \int d^3 \vec{R} G_1(\vec{R} - \Delta\vec{r}/2 , \vec{R} + \Delta\vec{r}/2)
    \label{eq: EqS4}
\end{equation}
where $S$ is the 2D trap volume, $\vec k= k_y \hat e_y+ k_z \hat{e_z}$ and $\vec\Delta r = \vec r- \vec r^\prime$.
From $\bar G_1(\Delta \vec{r})$ we determine the normalized first order correlation function $g_1(\vec \Delta r)= \bar G_1(\vec \Delta r)/ \bar G_1(0)$.

From our momentum distribution $n(\vec k)$ we extract the bidimensional first order correlation function, but as our turbulent cloud is anisotropic, we are interested in analyzing the direction with the largest momentum spread ($z$-axis). For this, we take the angular average of $g_1(\vec \Delta r)$ for the same angular regions that we compute $n(k,t)$  to obtain $g^{1D}_1(r)$. Due to a phase correlation component the first order correlation function is complex, in this sense we analyze its norm $|g^{1D}_1(r)|$ to determine the coherence length. 

The norm of the first order correlation function can be expressed as an Gaussian function \cite{bloch2000measurement} in the form
\begin{equation}
    |g_1(r)|=Be^{-r^2/2\ell^2} + c,
    \label{eq: EqS5}
\end{equation}
with $\ell$ being the coherence length, $B$ the initial amplitude of the decay and $c$ is an offset.  We determine the coherence length over the time evolution of our turbulent system $\ell(t)$, by fitting our experimentally determined $|g^{1D}_1(r)|$ to Eq.~(\ref{eq: EqS5}).
\\

\Sec{Simulation in a homogeneous potential}
Shaking the trap injects energy $\Delta E$ into the system, in addition to ground state energy $E_0$, triggering a direct turbulent cascade of energy that populates the high momentum region. After excitation stops, the direct cascade continues for a while as the low-momentum state gets depleted, leading to the NTFP stage. The final equilibrium state of a non-interacting or weakly-interacting boson gas at $k>0$ obeys the Bose-Einstein (BE) distribution \cite{pethick2008bose}: 
 \begin{equation} \label{eq: EqS6}
n^{\rm BE}(\epsilon_k) \;=\; \frac{1}{\exp\!\big[\frac{1}{k_B\,T}(\varepsilon_k-\mu)\big]-1},
 \end{equation}
where $\epsilon_k$ is the energy of a single particle, and the non-positive chemical potential $\mu$ and temperature $T$ are determined by the total energy $E=E_0+\Delta E$ and the number of particles $N$ of the closed system. Given a fixed $N$, whether there will be a condensate component in the final state at $k=0$  depends on $E$ compared to the critical value $E_c$ corresponding to the critical temperature $T_c$.  
If $E<E_c$, the BE distribution gives a pinned chemical potential, $\mu=0$,  yielding a macroscopic condensate
with a thermal tail. Dynamically, condensation proceeds via an inverse cascade: low-$k$ “blow-up” follows particle repopulation.
As $E\to E_c^{-}$, the condensate fraction decreases while the thermal fraction grows.
If $E>E_c$, the BE solution has a negative chemical potential, $\mu<0$, 
the condensate disappears, and the cloud equilibrates to a purely thermal gas. In this case, after the NTFP stage of the direct cascade, low-$k$ occupation is strongly depleted and no inverse cascade develops---there is no repopulation or condensation.
For an ideal gas in a uniform box, the critical temperature is fixed by the average particle density $n=N/V$ and atom mass $m$, $T_c=T_c(n,m)$ \cite{pethick2008bose, dalfovo1999theory}; for a cloud in a harmonic trap, $T_c=T_c(\bar\omega,N)$ with $\bar\omega=(\omega_x\omega_y\omega_z)^{1/3}$ \cite{ketterle1996bose}.
For a real (weakly interacting) gas, beyond-mean-field critical correlations raise $T_c$ in the uniform case (leading correction $\propto a n^{1/3}$ where $a$ is the $s$-wave scattering length) \cite{kashurnikov2001critical}, whereas in a harmonic trap  finite-$N$ and mean-field effects lower $T_c$ by small, known fractions \cite{ketterle1996bose, dalfovo1999theory}.

We consider a reduced uniform-gas model for the high-occupation sector ($n_k\gtrsim 1$), when $T<T_c$ or slightly greater than $T_c$. The BE distribution \eqref{eq: EqS6} is then well approximated by the Rayleigh–Jeans (RJ) law
\begin{equation} \label{eq: EqS7}
n_k=\frac{k_B T}{\hbar^2 k^2/(2m)-\mu}\,,
\end{equation}
with a momentum cutoff $k_c$, the gas is consequently described by a truncated Gross–Pitaevskii equation (modes $|\mathbf k|\le k_c$ only) \cite{krstulovic2011dispersive, kashurnikov2001critical}. We choose $k_c$ by imposing an edge occupation $n_k|_{k_c}=1$, which gives
$k_c(T,\mu)=\sqrt{\frac{2m}{\hbar^2}\Big(k_B T \ln 2+\mu\Big)}$. This cutoff in RJ distribution corresponds to a crossover $k$ in the BE distribution above which the RJ law changes into an exponentially decreasing (Maxwell-Boltzmann) dependence of the classical thermal component.
Near the transition we set $\mu\simeq 0$ and use
$k_c \equiv k_c(T_c,0)=\sqrt{\frac{2m}{\hbar^2}\,k_B T_c(\ln 2)}$,
treating $k_c$ as the boundary between low-energy coherent modes (kept) and high-energy thermal modes (omitted by this model).
The corresponding critical energy per particle of the truncated RJ gas is
\begin{equation}\label{eq: EqS8}
\left(\frac{E}{N}\right)_c=\frac{\hbar^2 k_c^2}{6m}
=\frac{\ln 2}{3}\,k_B T_c. 
\end{equation}

To demonstrate the thermalization process in both subcritical and supercritical scenarios, we perform numerical simulations of the dimensionless Gross-Pitaevskii equation,
\begin{equation}
    \frac{\partial \psi \left(\mathbf{x},t\right)}{\partial t} = i \left[\nabla^2 - | \psi \left(\mathbf{x},t\right)|^2\right] \psi \left(\mathbf{x},t\right),
    \label{eq: EqS9}
\end{equation}
within a triply periodic box of size $L=8\pi$, resulting in a volume $V=L^3$. This is achieved using the classic pseudospectral \cite{zhu2022,zhu2023direct} method with a $256^3$ grid truncated at $k_c=21$ \cite{zhu2023self}.
We perform  two simulations with the fixed total number of particles $N=10^5$, and Gaussian shaped initial distribution centering around a low and a
high momentum \cite{zhu2023self} ensuring that the energy falls below
and above $E_c$, respectively. The initial profiles can be
interpreted to be in the pre-thermalization stage. In this dimensionless framework, the condition \eqref{eq: EqS8} translates to $(E/N)_c=k_c^2/3$ \cite{connaughton2005condensation}.

\begin{figure}
\centering
    \includegraphics[scale=1]{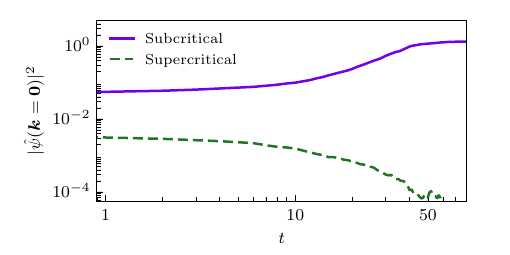}
    \caption{Time evolution of occupation number at zero momentum $|\hat\psi(\mathbf{k=0})|^2$. For the subcritical case we see the population of the particles at zero momentum. In the supercitical case we observe the depopulation of the low momenta.}
    \label{fig: FigS4}
\end{figure}

In Fig.~\ref{fig: FigS4} we can observe the behavior of the particles at zero momentum $|\hat\psi(\mathbf{k=0})|^2$. In the subcritical scenario an inverse cascade and a repopulation of particles at low momenta are clearly observed. In contrast, in the supercritical scenario  the low$-k$ region is directly thermalized. 

\bibliographystyle{apsrev4-2}
\bibliography{references.bib}

@article{
zhaoyu_2022,
author = {Zhao-Yu Zhou  and Guo-Xian Su  and Jad C. Halimeh  and Robert Ott  and Hui Sun  and Philipp Hauke  and Bing Yang  and Zhen-Sheng Yuan  and Jürgen Berges  and Jian-Wei Pan },
title = {Thermalization dynamics of a gauge theory on a quantum simulator},
journal = {Science},
volume = {377},
number = {6603},
pages = {311-314},
year = {2022},
doi = {10.1126/science.abl6277},
URL = {https://www.science.org/doi/abs/10.1126/science.abl6277},
}

@article{seman2011route,
  title={Route to turbulence in a trapped Bose-Einstein condensate},
  author={Seman, JA and Henn, Emanuel Alves de Lima and Shiozaki, RF and Roati, G and Poveda-Cuevas, FJ and Magalh{\~a}es, Kilvia Mayre Farias and Yukalov, VI and Tsubota, M and Kobayashi, M and Kasamatsu, K and others},
  journal={Laser Physics Letters},
  volume={8},
  number={9},
  pages={691--696},
  year={2011},
  publisher={Wiley Online Library},
  doi={10.1002/lapl.201110052},
  url={ https://doi.org/10.1002/lapl.201110052}
}

@article{madeira2020quantum,
  title={Quantum turbulence in quantum gases},
  author={Madeira, Lucas and Caracanhas, M{\^o}nica Andrioli and dos Santos, FEA and Bagnato, Vanderlei Salvador},
  journal={Annual Review of Condensed Matter Physics},
  volume={11},
  number={1},
  pages={37--56},
  year={2020},
  publisher={Annual Reviews},
  doi={10.1146/annurev-conmatphys-031119-050821},
  url={https://doi.org/10.1146/annurev-conmatphys-031119-050821}
}

@article{madeira2024universal,
author = {L. Madeira  and Arnol D. García-Orozco  and Michelle A. Moreno-Armijos  and Amilson R. Fritsch  and Vanderlei S. Bagnato },
title = {Universal scaling in far-from-equilibrium quantum systems: An equivalent differential approach},
journal = {Proceedings of the National Academy of Sciences},
volume = {121},
number = {30},
pages = {e2404828121},
year = {2024},
doi = {10.1073/pnas.2404828121},
URL = {https://www.pnas.org/doi/abs/10.1073/pnas.2404828121}
}

@article{garcia2022universal,
  title = {Universal dynamics of a turbulent superfluid {B}ose gas},
  author = {Garc\'{\i}a-Orozco, A. D. and Madeira, L. and Moreno-Armijos, M. A. and Fritsch, A. R. and Tavares, P. E. S. and Castilho, P. C. M. and Cidrim, A. and Roati, G. and Bagnato, V. S.},
  journal = {Physical Review A},
  volume = {106},
  issue = {2},
  pages = {023314},
  numpages = {10},
  year = {2022},
  month = {Aug},
  publisher = {American Physical Society},
  doi = {10.1103/PhysRevA.106.023314},
  url = {https://link.aps.org/doi/10.1103/PhysRevA.106.023314}
}

@article{moreno2025observation,
  title = {Observation of Relaxation Stages in a Nonequilibrium Closed Quantum System: {D}ecaying Turbulence in a Trapped Superfluid},
  author = {Moreno-Armijos, M. A. and Fritsch, A. R. and Garc\'{\i}a-Orozco, A. D. and Sab, S. and Telles, G. and Zhu, Y. and Madeira, L. and Nazarenko, S. and Yukalov, V. I. and Bagnato, V. S.},
  journal = {Physical Review Letters},
  volume = {134},
  issue = {2},
  pages = {023401},
  numpages = {6},
  year = {2025},
  month = {Jan},
  publisher = {American Physical Society},
  doi = {10.1103/PhysRevLett.134.023401},
  url = {https://link.aps.org/doi/10.1103/PhysRevLett.134.023401}
}

@article{navon2016emergence,
  title={Emergence of a turbulent cascade in a quantum gas},
  author={Navon, Nir and Gaunt, Alexander L and Smith, Robert P and Hadzibabic, Zoran},
  journal={Nature},
  volume={539},
  number={7627},
  pages={72-75},
  year={2016},
  publisher={Nature Publishing Group UK London},
  doi={10.1038/nature20114},
  url={https://doi.org/10.1038/nature20114}
}

@article{karailiev2024observation,
  title={Observation of an inverse turbulent-wave cascade in a driven quantum gas},
  author={Karailiev, Andrey and Gazo, Martin and Ga{\l}ka, Maciej and Eigen, Christoph and Satoor, Tanish and Hadzibabic, Zoran},
  journal={Physical Review Letters},
  volume={133},
  number={24},
  pages={243402},
  year={2024},
  publisher={APS},
 doi={10.1103/PhysRevLett.133.243402},
 url={https://link.aps.org/doi/10.1103/PhysRevLett.133.243402}
}

@article{pineiro2015universal,
  title={Universal self-similar dynamics of relativistic and nonrelativistic field theories near nonthermal fixed points},
  author={Pi{\~n}eiro Orioli, Asier and Boguslavski, Kirill and Berges, J{\"u}rgen},
  journal={Physical Review D},
  volume={92},
  number={2},
  pages={025041},
  year={2015},
  publisher={APS},
doi={10.1103/PhysRevD.92.025041},
url={https://doi.org/10.1103/PhysRevD.92.025041}
}

@article{Schmied2019,
author = {Schmied, Christian-Marcel and Mikheev, Aleksandr N. and Gasenzer, Thomas},
title = {Non-thermal fixed points: Universal dynamics far from equilibrium},
journal = {International Journal of Modern Physics A},
volume = {34},
number = {29},
pages = {1941006},
year = {2019},
doi = {10.1142/S0217751X19410069},
URL = {   
        https://doi.org/10.1142/S0217751X19410069
}
}

@article{mikheev2023universal,
  title={Universal dynamics and non-thermal fixed points in quantum fluids far from equilibrium},
  author={Mikheev, Aleksandr N and Siovitz, Ido and Gasenzer, Thomas},
  journal={The European Physical Journal Special Topics},
  volume={232},
  number={20},
  pages={3393--3415},
  year={2023},
  publisher={Springer},
doi={10.1140/epjs/s11734-023-00974-7},
url={https://doi.org/10.1140/epjs/s11734-023-00974-7
}
}

@article{zhu2023direct,
  title = {Direct and Inverse Cascades in Turbulent {B}ose-{E}instein Condensates},
  author = {Zhu, Ying and Semisalov, Boris and Krstulovic, Giorgio and Nazarenko, Sergey},
  journal = {Physical Review Letters},
  volume = {130},
  issue = {13},
  pages = {133001},
  numpages = {6},
  year = {2023},
  month = {Mar},
  publisher = {American Physical Society},
  doi = {10.1103/PhysRevLett.130.133001},
  url = {https://link.aps.org/doi/10.1103/PhysRevLett.130.133001}
}

@article{zhu2023self,
  title = {Self-similar evolution of wave turbulence in {G}ross-{P}itaevskii system},
  author = {Zhu, Ying and Semisalov, Boris and Krstulovic, Giorgio and Nazarenko, Sergey},
  journal = {Physical Review E},
  volume = {108},
  issue = {6},
  pages = {064207},
  numpages = {13},
  year = {2023},
  month = {Dec},
  publisher = {American Physical Society},
  doi = {10.1103/PhysRevE.108.064207},
  url = {https://link.aps.org/doi/10.1103/PhysRevE.108.064207}
}

@book{nazarenko2011wave,
  author    = {S. Nazarenko},
  title     = {Wave Turbulence},
  year      = {2011},
  publisher = {Springer},
  address   = {Berlin, Heidelberg},
  volume    = {825},
  series    = {Lecture Notes in Physics},
  isbn      = {978-3-642-15941-1},
  doi       = {10.1007/978-3-642-15942-8},
}

@book{zakharov1992kolmogorov,
  title     = {Kolmogorov Spectra of Turbulence I: Wave Turbulence},
  author    = {Vladimir E. Zakharov and Victor S. L'vov and Grigory Falkovich},
  year      = {1992},
  publisher = {Springer-Verlag},
  address   = {Berlin, Heidelberg},
  series    = {Springer Series in Nonlinear Dynamics},
  isbn      = {978-3-642-50052-7},
  doi       = {10.1007/978-3-642-50052-7},
}

@article{hohenberg1977theory,
  title = {Theory of dynamic critical phenomena},
  author = {Hohenberg, P. C. and Halperin, B. I.},
  journal = {Rev. Mod. Phys.},
  volume = {49},
  issue = {3},
  pages = {435--479},
  numpages = {0},
  year = {1977},
  month = {Jul},
  publisher = {American Physical Society},
  doi = {10.1103/RevModPhys.49.435},
  url = {https://link.aps.org/doi/10.1103/RevModPhys.49.435}
}

@article{dossantosfilho2022incompressible,
title = {Incompressible energy spectrum from wave turbulence},
journal = {Physica D: Nonlinear Phenomena},
volume = {440},
pages = {133479},
year = {2022},
issn = {0167-2789},
doi = {https://doi.org/10.1016/j.physd.2022.133479},
url = {https://www.sciencedirect.com/science/article/pii/S0167278922001981},
author = {{Dos Santos Filho, M. A. G.; Dos Santos, F. E. A.}}
}

@article{navon2019synthetic,
author = {Nir Navon  and Christoph Eigen  and Jinyi Zhang  and Raphael Lopes  and Alexander L. Gaunt  and Kazuya Fujimoto  and Makoto Tsubota  and Robert P. Smith  and Zoran Hadzibabic },
title = {Synthetic dissipation and cascade fluxes in a turbulent quantum gas},
journal = {Science},
volume = {366},
number = {6463},
pages = {382-385},
year = {2019},
doi = {10.1126/science.aau6103},
URL = {https://www.science.org/doi/abs/10.1126/science.aau6103}
}

@article{gatka2022emergence,
  title = {Emergence of Isotropy and Dynamic Scaling in 2D Wave Turbulence in a Homogeneous Bose Gas},
  author = {Ga\l{}ka, Maciej and Christodoulou, Panagiotis and Gazo, Martin and Karailiev, Andrey and Dogra, Nishant and Schmitt, Julian and Hadzibabic, Zoran},
  journal = {Phys. Rev. Lett.},
  volume = {129},
  issue = {19},
  pages = {190402},
  numpages = {6},
  year = {2022},
  month = {Nov},
  publisher = {American Physical Society},
  doi = {10.1103/PhysRevLett.129.190402},
  url = {https://link.aps.org/doi/10.1103/PhysRevLett.129.190402}
}

@article{kashurnikov2001critical,
  title={Critical temperature shift in weakly interacting Bose gas},
  author={Kashurnikov, VA and Prokof'ev, NV and Svistunov, BV},
  journal={Physical review letters},
  volume={87},
  number={12},
  pages={120402},
  year={2001},
  publisher={APS},
doi={10.1103/PhysRevLett.87.120402},
url={https://link.aps.org/doi/10.1103/PhysRevLett.87.120402}
}

@article{krstulovic2011dispersive,
  title={Dispersive bottleneck delaying thermalization of turbulent Bose-Einstein condensates},
  author={Krstulovic, Giorgio and Brachet, Marc},
  journal={Physical review letters},
  volume={106},
  number={11},
  pages={115303},
  year={2011},
  publisher={APS},
doi = {10.1103/PhysRevLett.106.115303},
  url = {https://link.aps.org/doi/10.1103/PhysRevLett.106.115303}
}

@article{connaughton2005condensation,
  title={Condensation of classical nonlinear waves},
  author={Connaughton, Colm and Josserand, Christophe and Picozzi, Antonio and Pomeau, Yves and Rica, Sergio},
  journal={Physical review letters},
  volume={95},
  number={26},
  pages={263901},
  year={2005},
  publisher={APS},
doi = {10.1088/1742-5468/2016/06/064009},
url = {https://doi.org/10.1088/1742-5468/2016/06/064009}
}

@article{langen2016prethermalization,
  title={Prethermalization and universal dynamics in near-integrable quantum systems},
  author={Langen, Tim and Gasenzer, Thomas and Schmiedmayer, J{\"o}rg},
  journal={Journal of Statistical Mechanics: Theory and Experiment},
  volume={2016},
  number={6},
  pages={064009},
  year={2016},
  publisher={IOP Publishing},
  doi={10.1088/1742-5468/2016/06/064009},
  url={https://doi.org/10.1088/1742-5468/2016/06/064009}
}

@article{Smith_2013,
doi = {10.1088/1367-2630/15/7/075011},
url = {https://doi.org/10.1088/1367-2630/15/7/075011},
year = {2013},
month = {jul},
publisher = {IOP Publishing},
volume = {15},
number = {7},
pages = {075011},
author = {Adu Smith, D and Gring, M and Langen, T and Kuhnert, M and Rauer, B and Geiger, R and Kitagawa, T and Mazets, I and Demler, E and Schmiedmayer, J},
title = {Prethermalization revealed by the relaxation dynamics of full distribution functions},
journal = {New Journal of Physics}
}

@article{chertkov2023characterizing,
  author       = {Eli Chertkov and Zihan Cheng and Andrew C. Potter and Sarang Gopalakrishnan and Thomas M. Gatterman and Justin A. Gerber and Kevin Gilmore and Dan Gresh and Alex Hall and Aaron Hankin and Mitchell Matheny and Tanner Mengle and David Hayes and Brian Neyenhuis and Russell Stutz and Michael Foss-Feig},
  title        = {Characterizing a non-equilibrium phase transition on a quantum computer},
  journal      = {Nature Physics},
  volume       = {19},
  number       = {12},
  pages        = {1799--1804},
  year         = {2023},
  doi          = {10.1038/s41567-023-02199-w},
  url          = {https://doi.org/10.1038/s41567-023-02199-w},
  issn         = {1745-2481}
}

@article{tatsuhiko2020general,
author = {Tatsuhiko N. Ikeda  and Masahiro Sato },
title = {General description for nonequilibrium steady states in periodically driven dissipative quantum systems},
journal = {Science Advances},
volume = {6},
number = {27},
pages = {eabb4019},
year = {2020},
doi = {10.1126/sciadv.abb4019},
URL = {https://www.science.org/doi/abs/10.1126/sciadv.abb4019}
}

@article{langen2015ultracold,
   author = "Langen, Tim and Geiger, Remi and Schmiedmayer, Jörg",
   title = "Ultracold Atoms Out of Equilibrium", 
   journal= "Annual Review of Condensed Matter Physics",
   year = "2015",
   volume = "6",
   number = "Volume 6, 2015",
   pages = "201-217",
   doi = "https://doi.org/10.1146/annurev-conmatphys-031214-014548",
   url = "https://www.annualreviews.org/content/journals/10.1146/annurev-conmatphys-031214-014548",
   publisher = "Annual Reviews",
   issn = "1947-5462",
   type = "Journal Article"
  }

@article{eisert2015quantum,
  author       = {J. Eisert and M. Friesdorf and C. Gogolin},
  title        = {Quantum many-body systems out of equilibrium},
  journal      = {Nature Physics},
  volume       = {11},
  number       = {2},
  pages        = {124--130},
  year         = {2015},
  doi          = {10.1038/nphys3215},
  url          = {https://doi.org/10.1038/nphys3215},
  issn         = {1745-2481}
}

@article{gazo2025science,
author = {Martin Gazo  and Andrey Karailiev  and Tanish Satoor  and Christoph Eigen  and Maciej Gałka  and Zoran Hadzibabic },
title = {Universal coarsening in a homogeneous two-dimensional Bose gas},
journal = {Science},
volume = {389},
number = {6762},
pages = {802-805},
year = {2025},
doi = {10.1126/science.ado3487},
URL = {https://www.science.org/doi/abs/10.1126/science.ado3487}}

@article{dalfovo1999theory,
  title = {Theory of Bose-Einstein condensation in trapped gases},
  author = {Dalfovo, Franco and Giorgini, Stefano and Pitaevskii, Lev P. and Stringari, Sandro},
  journal = {Rev. Mod. Phys.},
  volume = {71},
  issue = {3},
  pages = {463--512},
  numpages = {0},
  year = {1999},
  month = {Apr},
  publisher = {American Physical Society},
  doi = {10.1103/RevModPhys.71.463},
  url = {https://link.aps.org/doi/10.1103/RevModPhys.71.463}
}

@article{ketterle1996bose,
  title = {Bose-Einstein condensation of a finite number of particles trapped in one or three dimensions},
  author = {Ketterle, Wolfgang and van Druten, N. J.},
  journal = {Phys. Rev. A},
  volume = {54},
  issue = {1},
  pages = {656--660},
  numpages = {0},
  year = {1996},
  month = {Jul},
  publisher = {American Physical Society},
  doi = {10.1103/PhysRevA.54.656},
  url = {https://link.aps.org/doi/10.1103/PhysRevA.54.656}
}

@article{zhu2022,
  title = {Testing wave turbulence theory for the Gross-Pitaevskii system},
  author = {Zhu, Ying and Semisalov, Boris and Krstulovic, Giorgio and Nazarenko, Sergey},
  journal = {Phys. Rev. E},
  volume = {106},
  issue = {1},
  pages = {014205},
  numpages = {18},
  year = {2022},
  month = {Jul},
  publisher = {American Physical Society},
  doi = {10.1103/PhysRevE.106.014205},
  url = {https://link.aps.org/doi/10.1103/PhysRevE.106.014205}
}

@article{castin_dum1996,
  title = {Bose-Einstein Condensates in Time Dependent Traps},
  author = {Castin, Y. and Dum, R.},
  journal = {Phys. Rev. Lett.},
  volume = {77},
  issue = {27},
  pages = {5315--5319},
  numpages = {0},
  year = {1996},
  month = {Dec},
  publisher = {American Physical Society},
  doi = {10.1103/PhysRevLett.77.5315},
  url = {https://link.aps.org/doi/10.1103/PhysRevLett.77.5315}
}

@article{Holly2023,
  title = {Strong quantum turbulence in Bose-Einstein condensates},
  author = {Middleton-Spencer, H. A. J. and Orozco, A. D. G. and Galantucci, L. and Moreno, M. and Parker, N. G. and Machado, L. A. and Bagnato, V. S. and Barenghi, C. F.},
  journal = {Phys. Rev. Res.},
  volume = {5},
  issue = {4},
  pages = {043081},
  numpages = {10},
  year = {2023},
  month = {Oct},
  publisher = {American Physical Society},
  doi = {10.1103/PhysRevResearch.5.043081},
  url = {https://link.aps.org/doi/10.1103/PhysRevResearch.5.043081}
}

@article{Henn2009,
  title = {Emergence of Turbulence in an Oscillating Bose-Einstein Condensate},
  author = {Henn, E. A. L. and Seman, J. A. and Roati, G. and Magalh\~aes, K. M. F. and Bagnato, V. S.},
  journal = {Phys. Rev. Lett.},
  volume = {103},
  issue = {4},
  pages = {045301},
  numpages = {4},
  year = {2009},
  month = {Jul},
  publisher = {American Physical Society},
  doi = {10.1103/PhysRevLett.103.045301},
  url = {https://link.aps.org/doi/10.1103/PhysRevLett.103.045301}
}

@article{bloch2000measurement,
  title={Measurement of the spatial coherence of a trapped Bose gas at the phase transition},
  author={Bloch, Immanuel and H{\"a}nsch, Theodor W and Esslinger, Tilman},
  journal={Nature},
  volume={403},
  number={6766},
  pages={166--170},
  year={2000},
  publisher={Nature Publishing Group UK London},
  doi = {10.1038/35003132},
  url = {https://doi.org/10.1038/35003132}
}

@article{naraschewski1999,
  title = {Spatial coherence and density correlations of trapped Bose gases},
  author = {Naraschewski, M. and Glauber, R. J.},
  journal = {Phys. Rev. A},
  volume = {59},
  issue = {6},
  pages = {4595--4607},
  numpages = {0},
  year = {1999},
  month = {Jun},
  publisher = {American Physical Society},
  doi = {10.1103/PhysRevA.59.4595},
  url = {https://link.aps.org/doi/10.1103/PhysRevA.59.4595}
}

@article{Bray01061994,
author = {A.J. Bray},
title = {Theory of phase-ordering kinetics},
journal = {Advances in Physics},
volume = {43},
number = {3},
pages = {357--459},
year = {1994},
publisher = {Taylor \& Francis},
doi = {10.1080/00018739400101505},
URL = { 
     https://doi.org/10.1080/00018739400101505

}
}

@article{martirosyan2025universal,
  title={A universal speed limit for spreading of coherence},
  author={Martirosyan, Gevorg and Gazo, Martin and Etrych, Ji{\v{r}}{\'\i} and Fischer, Simon M and Morris, Sebastian J and Ho, Christopher J and Eigen, Christoph and Hadzibabic, Zoran},
  journal={Nature},
  pages={1--5},
  year={2025},
  publisher={Nature Publishing Group UK London},
  doi={10.1038/s41586-025-09735-z},
  url={https://doi.org/10.1038/s41586-025-09735-z}
}

@misc{Temp_u0,
  note = {In this sense, we have analyzed the temperature of the system for long holding times, $t\approx400$ ms, for different excitation amplitudes calibrated in terms of the chemical potential $\mu_0$ of our BEC.}
}

@misc{SupMat,
      note={See Supplemental Material at [URL will be inserted by publisher] for more details on the experimental method, measurement of the momentum distributions, simulations in a homogeneous potential and the determination of the coherence lentgh.} }

@misc{TempAprox,
  author = {---},
  title = {},
  howpublished = {The temperature is obtained using a bimodal function fitting -- the combination of Thomas-Fermi and Gaussian profiles. This estimation does not accurately describe the far-from-equilibrium cloud during all the relaxation processes, so we refer to it as an approximated temperature.}
}

@article{PhysRevA.59.4595,
  title = {Spatial coherence and density correlations of trapped Bose gases},
  author = {Naraschewski, M. and Glauber, R. J.},
  journal = {Phys. Rev. A},
  volume = {59},
  issue = {6},
  pages = {4595--4607},
  numpages = {0},
  year = {1999},
  month = {Jun},
  publisher = {American Physical Society},
  doi = {10.1103/PhysRevA.59.4595},
  url = {https://link.aps.org/doi/10.1103/PhysRevA.59.4595}
}

@book{pethick2008bose,
address={Cambridge},
edition={2},
title={Bose–{E}instein Condensation in dilute gases}, publisher={Cambridge University Press},
author={Pethick, C. J. and Smith, H.},
year={2008},
pages={},
isbn={978-0521846516}}

\end{document}